\documentclass[12pt,letterpaper]{article}
\usepackage[utf8]{inputenc}

\usepackage{url,hyperref}
\usepackage{authblk}
\usepackage{paralist,enumitem}
\usepackage{graphicx}
\usepackage{color}

\title{Epidemiological data challenges: planning for a more robust future through data standards}
\author[1]{Geoffrey Fairchild\thanks{gfairchild@lanl.gov}}
\author[1]{Byron Tasseff}
\author[1]{Hari Khalsa}
\author[2]{Nicholas Generous}
\author[1]{Ashlynn R. Daughton}
\author[3]{Nileena Velappan}
\author[4]{Reid Priedhorsky}
\author[3]{Alina Deshpande}
\affil[1]{Analytics, Intelligence, and Technology Division, Los Alamos National Laboratory, Los Alamos, New Mexico, USA}
\affil[2]{Intelligence and Emerging Threats Program Office, Los Alamos National Laboratory, Los Alamos, New Mexico, USA}
\affil[3]{Bioscience Division, Los Alamos National Laboratory, Los Alamos, New Mexico, USA}
\affil[4]{High Performance Computing Division, Los Alamos National Laboratory, Los Alamos, New Mexico, USA}
\date{2018-11-14}

\begin{document}

\begin{titlepage}
\maketitle
\end{titlepage}

\begin{abstract}

Accessible epidemiological data are of great value for emergency preparedness and response, understanding disease progression through a population, and building statistical and mechanistic disease models that enable forecasting. The status quo, however, renders acquiring and using such data difficult in practice. In many cases, a primary way of obtaining epidemiological data is through the internet, but the methods by which the data are presented to the public often differ drastically among institutions. As a result, there is a strong need for better data sharing practices. This paper identifies, in detail and with examples, the three key challenges one encounters when attempting to acquire and use epidemiological data:
\begin{inparaenum}[1)]
    \item \emph{interfaces},
    \item \emph{data formatting}, and
    \item \emph{reporting}.
\end{inparaenum}
These challenges are used to provide suggestions and guidance for improvement as these systems evolve in the future. If these suggested data and interface recommendations were adhered to, epidemiological and public health analysis, modeling, and informatics work would be significantly streamlined, which can in turn yield better public health decision-making capabilities.

\bigskip

\end{abstract}
\section{Introduction}

At the heart of disease surveillance and modeling are \emph{epidemiological data}. These data are generally presented as a time series of cases, $T$, for a geographic region, $G$, and for a demographic, $D$. The type of cases presented may vary depending on the context. For example, $T$ may be a time series of confirmed or suspected cases, or it might be hospitalizations or deaths; in some circumstances, it may be a summation of some combination of these (e.g., confirmed $+$ suspected cases). $G$ is most commonly a political boundary; it might be a country, state/province, county/district, city, or sub-city region, such as a postal code or United States (U.S.) Census Bureau census tract. Depending on the context, $D$ may simply be the the entire population of $G$, or it might be stratified by age, sex, race, education, or other relevant factors.

Epidemiological data have a variety of uses. From a public health perspective, they can be used to gain an understanding of population-level disease progression. This understanding can in turn be used to aid in decision-making and allocation of resources. Recent outbreaks like Ebola and Zika have demonstrated the value of accessible epidemiological data for emergency preparedness and the need for better data sharing~\cite{Chretien2016}. These data may influence vaccine distribution~\cite{U.S.DepartmentofHealthandHumanServices2008b}, and hospitals can anticipate surge capacity during an outbreak, allowing them to obtain extra temporary help if necessary~\cite{Nap2007, Hota2010}.

From a modeler's perspective, high quality reference data (also commonly referred to as ground truth data) are needed to enable prediction and forecasting~\cite{Moran2016}. These data can be used to parameterize compartmental models~\cite{Hethcote2000} as well as stochastic agent-based models (e.g., \cite{Eubank2004, Bisset2009, Chao2012, Grefenstette2013, McMahon2014}). They can also be used to train and validate machine learning and statistical models (e.g., \cite{Viboud2003, Polgreen2008, Ginsberg2009, Signorini2011, Shaman2013, Generous2014, Hickmann2015, Fairchild2015}).

The internet has become the predominant way to publish, share, and collect epidemiological data. While data standards exist for observational studies~\cite{STROBE2018} and clinical research~\cite{CDISC2018}, for example, no such standards exist for the publication of the kind of public health-related epidemiological data described above. Despite the strong need to share and consume data, there are many legal, technical, political, and cultural challenges in implementing a standardized epidemiological data framework~\cite{Pisani2010a, Sane2015a}. As a result, the methods by which data are presented to the public often differ significantly among data-sharing institutions (e.g., public health departments, ministries of health, data collection or aggregation services). Moreover, these problems are not unique to epidemiological data; the issues described in this paper are common across many different disciplines.

First, epidemiological data on the internet are presented to the user through a variety of \emph{interfaces}. These interfaces vary widely not only in their appearance but also in their functionality. Some data are openly available through clear modern web interfaces, complete with well-documented programmer-friendly application programming interfaces (APIs), while others are displayed as static web pages that require error-prone and brittle web scraping. Still others are offered as machine-readable documents (e.g., comma-separate values (CSV), Microsoft Excel, Extensible Markup Language (XML), Adobe PDF). Finally, some necessitate contacting a human, who then prepares and sends the requested data manually.

Second, there are many \emph{data formats}. Data containers (e.g., CSV, JavaScript Object Notation (JSON)) and element formats (e.g., timestamp format, location name format) may differ. Character encodings~\cite{Zentgraf2015} (e.g., ASCII, UTF-8) and line endings~\cite{Atwood2010} (e.g., \texttt{\textbackslash r\textbackslash n}, \texttt{\textbackslash n}) may also differ. Compounding these issues, formats can change over time (e.g., renaming or reordering spreadsheet columns). More broadly, these challenges are closely tied to schema, data model, and vocabulary standardization.

Finally, there are differences among institutions in their \emph{reporting} habits; even within a single institution, there are often reporting nuances among diseases. For example, one context may be reported monthly (e.g., Q fever in Australia), while another context is reported weekly (e.g., influenza in the U.S.) or even more finely (e.g., 2014 West African Ebola outbreak). Furthermore, what is meant by ``weekly'' in one context may be different than another context (e.g., CDC epi weeks vs. irregular reporting intervals in Poland, as described later).

Together, these challenges make large-scale public health data analysis and modeling significantly more difficult and time-consuming. Gathering, cleaning, and eliciting relevant data often require more time than the actual analysis itself. This paper discusses these three key technical challenges involving public health-related epidemiological data, in detail and with examples that were identified through detailed analysis of data deposition practices around the globe. Building from this analysis, we offer a framework of best practices comprised of modern standards that should be adhered to when releasing epidemiological data to the public. Such a framework will enable a more robust future for accurate and high-confidence epidemiological data and analysis.
\section{Discussion}

\subsection{Interface challenges}

The \emph{interface} is the mechanism by which data are presented to a user for consumption.

Epidemiological data repositories implementing current best practices provide an interactive web-based searching and filtering interface that enables users to easily export desired data in a variety of formats. These are generally accompanied by an API that allows users to programmatically acquire desired data. For example, if one wants to download the latest influenza surveillance data weekly, instead of manually navigating an interactive web interface each week to export the data, the process could be automated by writing code that interacts with the API. Such an interface provides the simplest and most powerful method of data acquisition. Examples of this type of interface are the U.S. Centers for Disease Control and Prevention (CDC)\footnote{\url{https://data.cdc.gov/}} and the World Health Organization (WHO) Global Health Observatory (GHO)\footnote{\url{http://www.who.int/gho/en/}}.

While an interactive web-based interface coupled with an API is a best practice, it can be complex and expensive to implement. Many public health departments are under resource constraints and depend on older websites that tend to release data in one of two ways:
\begin{inparaenum}[1)]
    \item data are uploaded in some common format (e.g., CSV, Microsoft Excel, PDF) or
    \item data are displayed in Hypertext Markup Language (HTML) tables.
\end{inparaenum}
An example of the first is seen via Israel's Ministry of Health website, where data are provided weekly in Microsoft Excel formats~\cite{israel_web}. An example of the second is seen via Australia's Department of Health website, where data are provided within simply-formatted HTML tables~\cite{australia_web}.

Data uploaded in a common format can often be automatically downloaded and processed, and HTML tables can generally be automatically scraped and processed. While HTML scraping is often straightforward, there are some instances where it can be quite difficult. One example of a difficult-to-scrape data source is the Robert Koch Institute SurvStat 2.0 website~\cite{germany_web}. Although the service is capable of providing epidemiological data at superior spatial and temporal resolutions (county- and week-level, respectively), the interface is not easily amenable to scraping. First, the HTTP requests formed by the ASP.NET application cannot be easily reverse-engineered; this necessitates the use of browser-automation software like Selenium\footnote{\url{http://www.seleniumhq.org/}}, which enables automating website user interaction, such as mouse clicks and keyboard presses, for data scraping. Second, the selection of new filters, attributes, and display options results in a newly-refreshed page for each change; because many options are required to obtain each desired dataset, scraping can take a long time.

Additionally, while there may be no technical barriers to downloading or scraping data, there may be barriers relating to a website's terms of service (TOS). In some instances, the TOS may prevent users from scraping or downloading data en masse; this is sometimes done to prevent unreasonable load on the website, for example. Ignoring the TOS raises ethical issues that are often overlooked in research; after all, the goals of most epidemiological researchers are benevolent, and the data are public and usually funded by taxpayers. Ignoring a website's TOS could also raise logistical issues related to publishing and institutional review board (IRB) approval.

A concern underlying all scraping efforts is that data scraping scripts are brittle. Web scraping relies on patterns in the HTML/CSS source code of a website. If an institution modifies its layouts, even slightly, scrapers may exhibit unexpected behavior.

In some cases, a human must be contacted directly, who then prepares and sends the requested data. However, these manually requested and prepared data are often saddled with many restrictions. For example, when one of the authors contacted a ministry of health for more detailed epidemiological data, the data were offered with a five-page data request form that significantly restricted use and sharing of the data. Furthermore, it stated that it would take ``up to three months'' to be released because of the review and approval from the various data owners (local, state, and territory health departments). These types of restrictions and hurdles to data access prevent the development and adoption of advanced analytics.

\begin{figure}[!ht]
    \centering
    \includegraphics[width=\textwidth]{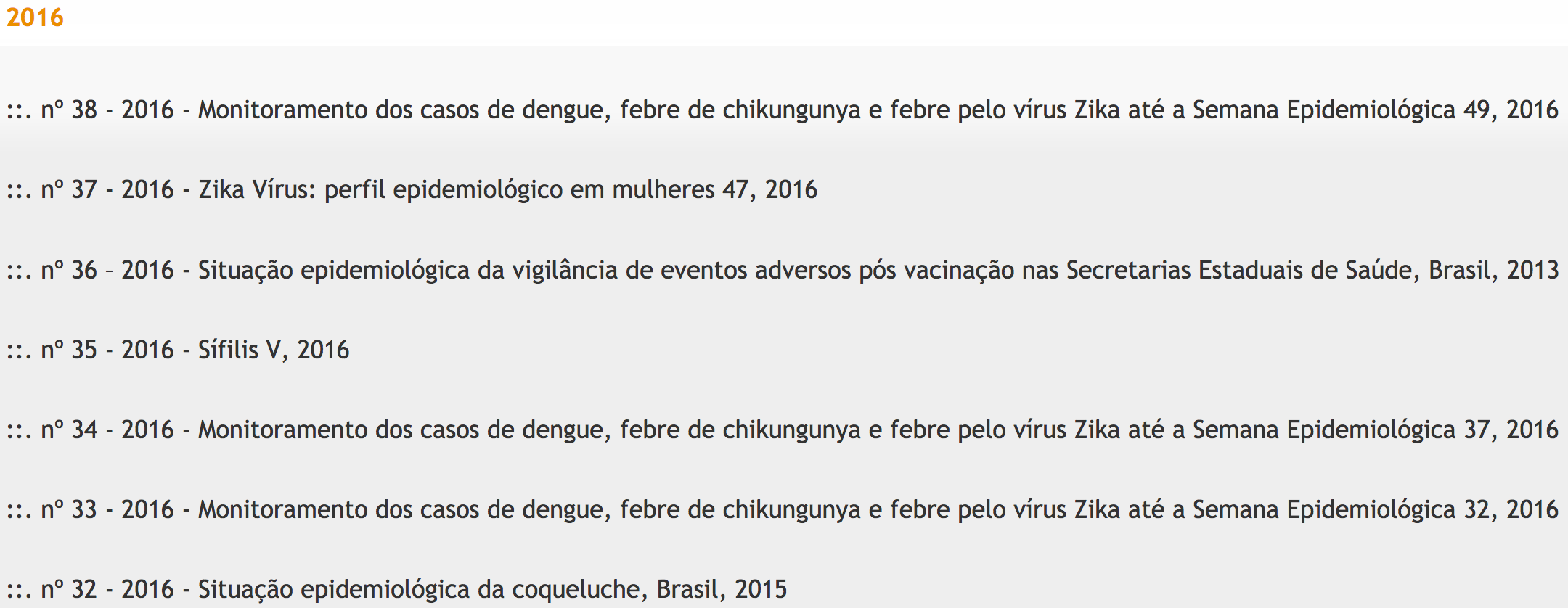}
    \caption{\textbf{Screenshot showing part of the mosquito-borne illness epidemiological bulletin list available at~\cite{brazil_zika_list}.} This is the most current and complete list, with data available through the 38th week of 2016.}
    \label{figure:brazil_mosquito_bulletins_current}
\end{figure}

\begin{figure}[!ht]
    \centering
    \includegraphics[width=\textwidth]{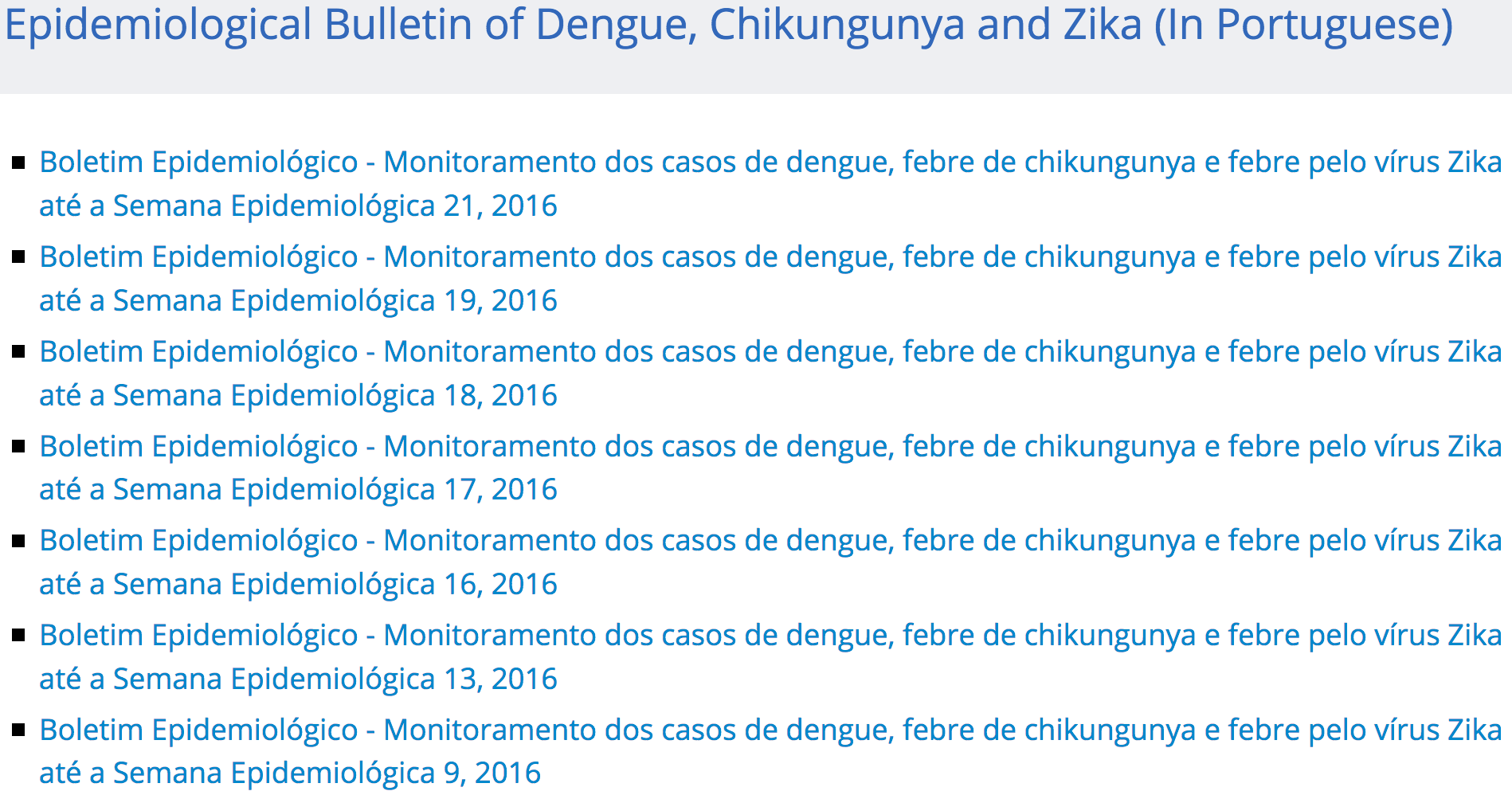}
    \caption{\textbf{Screenshot showing part of the mosquito-borne illness epidemiological bulletin list available at~\cite{brazil_english_zika_list}.} This list only goes through week 21 of 2016 and is missing a number of weeks when compared to the list in Figure~\ref{figure:brazil_mosquito_bulletins_current}. This screenshot was taken at the same time as the one in Figure~\ref{figure:brazil_mosquito_bulletins_current}.}
    \label{figure:brazil_mosquito_bulletins_current_old}
\end{figure}

Finally, finding epidemiological data interfaces or data within an interface is often a time-consuming and error-prone task. For example, the Zika virus epidemic has resulted in increased global attention for Brazil, but it has not resulted in a single easy-to-understand machine-readable interface~\cite{Coelho2016}. Until just recently, Brazil's Ministry of Health maintained two separate lists of mosquito-borne illness epidemiological bulletins~\cite{brazil_english_zika_list, brazil_zika_list}. Although these lists pointed to the exact same bulletins, \cite{brazil_zika_list} is consistently more up-to-date than~\cite{brazil_english_zika_list} (see Figures~\ref{figure:brazil_mosquito_bulletins_current} and \ref{figure:brazil_mosquito_bulletins_current_old}). Having multiple interfaces increases the likelihood of human error when collecting epidemiological data. For instance, if one assumes that there is only one official source for Zika, the most current information may be overlooked.

\subsection{Data format challenges}

The \emph{data format} specifies how the data are read and written. There are two layers:
\begin{inparaenum}[1)]
    \item the \emph{data container} and
    \item the \emph{element format}.
\end{inparaenum}
The data container specifies how individual elements should be agglomerated; CSV is an example of a data container. The element format specifies how each individual element should be arranged; the ISO 8601 date and time specification is an example of an element formatting standard.

Data format challenges often provide the biggest obstacles that users must overcome. In order for an analyst to use data from multiple sources, they must first be merged. In practice, however, data from one institution are seldom available in a format that can be directly compared to data from another institution.

\subsubsection*{Data containers}

First, data container issues must be addressed. For example, CSV files are among the simplest file types to parse; they are plain text files with a simple structure (i.e., columns are separated by a comma, rows are separated by a newline). Figure~\ref{figure:csv_example} demonstrates how epidemiological data might be provided in a CSV file. Any spreadsheet software can open CSV files natively, and most programming languages require no third-party libraries to read and write CSV files.

\begin{figure}[!ht]
    \centering
    \begin{verbatim}date,location,cases
2013-11-05,United States,4
2013-11-05,Germany,8
2013-11-11,South Africa,9
2013-11-12,Japan,6\end{verbatim}
    \caption{\textbf{Sample epidemiological case count data in CSV format.} CSV files are plain text files that allow tabular data to be laid out as rows separated by newlines and columns separated by commas. This time series does not contain real data and only exists for demonstration purposes.}
    \label{figure:csv_example}
\end{figure}

A conceptually similar file type to CSV is Microsoft Excel's XLSX. XLSX is a spreadsheet format developed by Microsoft and is a part of the Office Open XML (OOXML) specification. OOXML is a complex specification comprised of zipped XML files and other embedded data (e.g., images)~\cite{OpenOffice2016}. This format is common among public health practitioners due to the ubiquity of Microsoft Excel. For the programmer, however, this format presents a variety of challenges not present with CSV files. Due to the file type's complexity, a third-party library will be necessary in virtually all circumstances for reading/writing XLSX files (e.g., xlrd\footnote{\url{https://github.com/python-excel/xlrd}} and xlwt\footnote{\url{https://github.com/python-excel/xlwt}} for Python, Apache POI\footnote{\url{https://poi.apache.org/}} for Java). Depending on the maturity of the library used, formulas, pivot tables, and other complex features should be handled with varying degrees of trust.

JSON is another common data container used on the internet (see Figure~\ref{figure:json_example} for an example). For instance, JSON data are commonly returned when querying an API endpoint. JSON is easy and fast to use; many programming languages offer built-in JSON read/write support (e.g., Python and Java). Additionally, similar to CSV, JSON is a plain text format that is human-readable. Unlike CSV, however, JSON is not limited to tabular data. JSON can represent more complex relationships between data and is conceptually more similar to XML. Due to its ubiquity and structure, a number of application-specific JSON standards are available. For example, GeoJSON~\cite{GeoJSON2016} and TopoJSON~\cite{TopoJSON2016} enable sharing geographic data. In 2016, Finnie et al. proposed EpiJSON, which offers a standardized way to encode epidemiological data~\cite{Finnie2016}. Although EpiJSON shows promise, it is young and has yet to be broadly embraced. To make adoption simpler, open-source EpiJSON libraries could be developed for common programming languages; currently, the only such library exists for the programming language R~\cite{repijson2015}, but additional libraries should be developed for Python, Java, and other languages commonly used for epidemiological data analysis.

\begin{figure}[!ht]
    \centering
    \begin{verbatim}[
  {
    "date":"2013-11-05",
    "locations":{
      "United States":4,
      "Germany":8
    }
  },
  {
    "date":"2013-11-11",
    "locations":{
      "South Africa":9
    }
  },
  {
    "date":"2013-11-12",
    "locations":{
      "Japan":6
    }
  }
]\end{verbatim}
    \caption{\textbf{Sample epidemiological case count data in a simple JSON format.} Compared to CSV (demonstrated in Figure~\ref{figure:csv_example}), JSON contains more structure that can more rigorously specify data relationships (including hierarchical relationships). Note that this is \emph{not} EpiJSON; EpiJSON can be quite verbose (due to, for example, metadata specifications and GeoJSON-specified locations), and the authors felt a complete EpiJSON example would take up an unreasonable amount of space in this paper. As in Figure~\ref{figure:csv_example}, this time series does not contain real data and only exists for demonstration purposes.}
    \label{figure:json_example}
\end{figure}

PDF files provide a number of unique challenges in addition to complexity. Extraction of data is the biggest challenge, as epidemiological information is often provided in mixed formats: textual (e.g., paragraphs of descriptive text in a report), graphical (e.g., bar and line charts), and tabular. Simply extracting the text of a PDF correctly and in the right order can prove to be a non-trivial challenge. Named-entity recognition and extraction, a natural language processing task, can be used to elicit case counts from unstructured text~\cite{Fairchild2015}, but this supervised machine learning task requires knowledge of the language in which the document is published, as well as epidemiological subject matter expertise. Graphical data are intended for the human eye. While graphical data can potentially be digitized using software like WebPlotDigitizer\footnote{\url{http://arohatgi.info/WebPlotDigitizer/}}, this cannot always be reliably automated. Even tabular data, which visually appear structured, are typically difficult to extract due to the variety of ways a table can be presented in a PDF document~\cite{Khusro2015}. 

Furthermore, PDF files need not even contain text. In a number of circumstances, the PDF files that institutions provide simply contain scanned images of documents. The resulting PDF simply contains the image, rather than the raw text that comprised the original document. For example, many of the weekly reports available through the Department of Health website for the Philippines~\cite{PhilippinesDOH2016} are PDFs of scanned documents (e.g., \cite{PhilippinesDOHDengue2015, PhilippinesDOHDiptheria2016}). Text can potentially be elicited with optical character recognition (OCR) software, but the quality of the resulting textual data will vary significantly depending on the quality of the scanned images.

Finally, one must be aware of the character encoding when reading text. Since, at the basic level, computers represent all data using binary bits, there must be some binary representation of each character or symbol in an alphabet or language; the character encoding specifies how the raw bits stored in a file should be converted to readable text and vice versa~\cite{Zentgraf2015}. While there are a number of possible encodings, ASCII, ISO-8859-1, and UTF-8 are among the most common encodings encountered in practice. In 2012, UTF-8 surpassed 60\% adoption across the web~\cite{Davis2012} and is currently approaching the 90\% mark~\cite{W3Techs2016}. Encoding differences are important; for example, while reading ASCII text as UTF-8 yields correct results, the converse does not.

\subsubsection*{Element format}

\paragraph*{Date and time}

Beyond data container challenges, there are a number of element formatting differences that must be addressed. First and foremost are date and time formatting discrepancies. While the ISO 8601 date/time standard has existed since 1988, it is often bypassed in favor of locale-dependent formats. For example, much of Europe follows the \texttt{day-month-year} convention, the U.S. follows the \texttt{month-day-year} convention, and China follows the \texttt{year-month-day} convention. Depending on the locale, \texttt{03-09-2005} may refer to March 9, 2005 or September 3, 2005. Additionally, not all locales use the Gregorian calendar. Thailand, for example, uses the Buddhist calendar. The current year, as represented by the Gregorian calendar, is 2018; a Thai timestamp would instead specify 2561. Finally, some locales use 24-hour time, while others use 12-hour time.

In addition to these timestamp parsing differences, there are significant \emph{implicit} timestamp differences that must be understood. To understand these, one must first recognize that a timestamp on a typical disease curve usually implicitly refers to an interval of time (i.e., actual event-level epidemiological data are rare). To illustrate this, consider the time series in Table~\ref{table:sample_time_series}. Each timestamp can be interpreted using one of three possible \emph{interval types}:
\begin{description}
    \item[Leading] The timestamp starts the interval, and the interval ends the ``instant'' before the next specified timestamp. Table~\ref{table:interval_type_leading} shows how the time series in Table~\ref{table:sample_time_series} would be transformed to an \emph{interval series} with an interval type of leading.
    \item[Trailing exclusive] The timestamp ends the interval but is not included in the interval; Table~\ref{table:interval_type_trailing_exclusive} demonstrates this transformation.
    \item[Trailing inclusive] The timestamp ends the interval and is included in the interval; Table~\ref{table:interval_type_trailing_inclusive} shows this transformation.
\end{description}
Note that we do not currently feel it is necessary to include a leading exclusive option; leading will always be inclusive.

\begin{table}[!ht]
    \centering
    \caption{\textbf{Sample historical weekly epidemiological time series consisting of timestamps and case counts.}}
    \label{table:sample_time_series}
    \begin{tabular}{c | c}
        \textbf{Timestamp} & \textbf{Cases} \\
        \hline
        \texttt{2014-08-07 00:00} & 2 \\
        \texttt{2014-08-14 00:00} & 5 \\
        \texttt{2014-08-21 00:00} & 4
    \end{tabular}
\end{table}

\begin{table}[!ht]
    \centering
    \caption{\textbf{Explicit transformation of Table~\ref{table:sample_time_series} into a \emph{leading} interval series.} The interval start and end are inclusive and exclusive, respectively.}
    \label{table:interval_type_leading}
    \begin{tabular}{c | c | c}
        \textbf{Interval start} & \textbf{Interval end} & \textbf{Cases} \\
        \hline
        \texttt{2014-08-07 00:00} & \texttt{2014-08-14 00:00} & 2 \\
        \texttt{2014-08-14 00:00} & \texttt{2014-08-21 00:00} & 5 \\
        \texttt{2014-08-21 00:00} & \texttt{2014-08-28 00:00} & 4
    \end{tabular}
\end{table}
    
\begin{table}[!ht]
    \centering
    \caption{\textbf{Explicit transformation of Table~\ref{table:sample_time_series} into a \emph{trailing exclusive} interval series.} The interval start and end are inclusive and exclusive, respectively.}
    \label{table:interval_type_trailing_exclusive}
    \begin{tabular}{c | c | c}
        \textbf{Interval start} & \textbf{Interval end} & \textbf{Cases} \\
        \hline
        \texttt{2014-07-31 00:00} & \texttt{2014-08-07 00:00} & 2 \\
        \texttt{2014-08-07 00:00} & \texttt{2014-08-14 00:00} & 5 \\
        \texttt{2014-08-14 00:00} & \texttt{2014-08-21 00:00} & 4
    \end{tabular}
\end{table}
    
\begin{table}[!ht]
    \centering
    \caption{\textbf{Explicit transformation of Table~\ref{table:sample_time_series} into a \emph{trailing inclusive} interval series.} The interval start and end are inclusive and exclusive, respectively.}
    \label{table:interval_type_trailing_inclusive}
    \begin{tabular}{c | c | c}
        \textbf{Interval start} & \textbf{Interval end} & \textbf{Cases} \\
        \hline
        \texttt{2014-08-01 00:00} & \texttt{2014-08-08 00:00} & 2 \\
        \texttt{2014-08-08 00:00} & \texttt{2014-08-15 00:00} & 5 \\
        \texttt{2014-08-15 00:00} & \texttt{2014-08-22 00:00} & 4
    \end{tabular}
\end{table}

\begin{figure}[!ht]
    \centering
    \includegraphics[width=\textwidth]{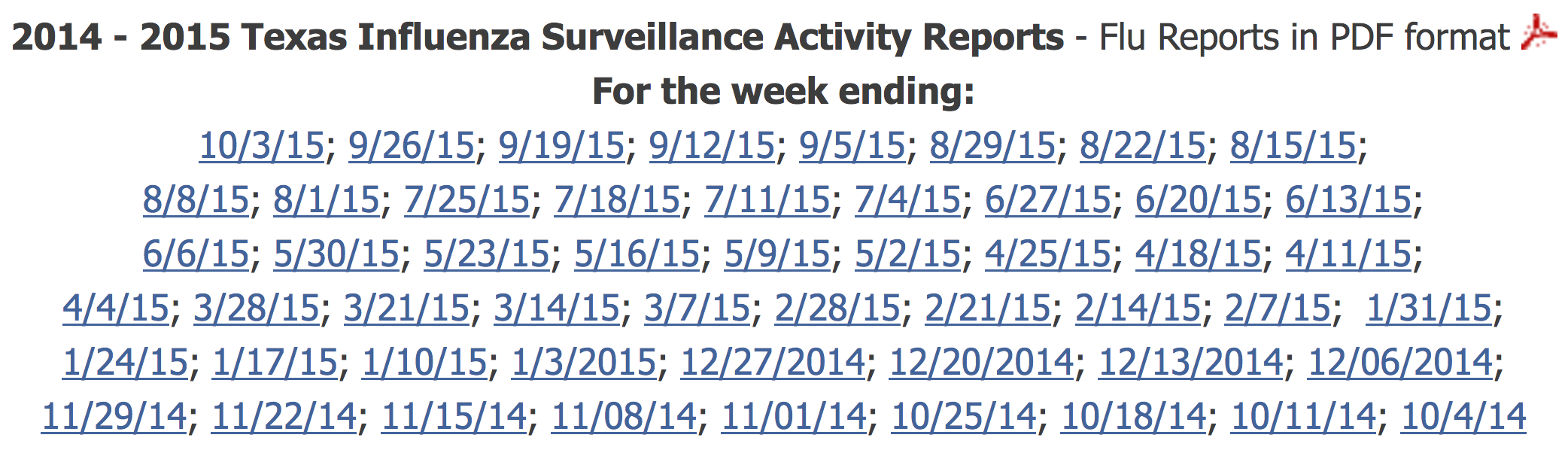}
    \caption{\textbf{Screenshot taken from Texas' Department of State Health Services 2014--2015 weekly influenza reports web page~\cite{1415FluTexasDoSHS2016}.} Texas identifies its weekly influenza surveillance PDF reports by trailing inclusive timestamps rather than by MMWR week (e.g., ``10/3/15'' instead of ``2015, week 39''). Interestingly, much of the data in each PDF uses MMWR week numbers rather than timestamps.}
    \label{figure:texas}
\end{figure}

As an example, the CDC standardizes reporting dates in the U.S. using the notion of an ``MMWR week'' or ``epi week''~\cite{CDC2016}. MMWR weeks always begin on Sundays and end on Saturdays. Weeks can be numbered 1--53, and, as a result, many institutions choose to report them as such (e.g., the interval [\texttt{2016-05-15 00:00}, \texttt{2016-05-22 00:00}) is reported as ``2016, week 20''). However, while most U.S.-based health departments respect the weekly MMWR aggregation standard, many continue to report timestamps based on the MMWR week concept. For example, Figure~\ref{figure:texas} shows how Texas identifies its weekly influenza surveillance PDF reports by trailing inclusive timestamps rather than by MMWR week.

Outside of the U.S., a variety of reporting date standards exist. In Japan, for example, the epi week starts on Monday and ends on Sunday~\cite{JapanEpiWeek2016}. In Poland, the reporting is even more different. Poland reports influenza cases four ``weeks'' a month, regardless of the length of the month. As a result, the intervals between reports are not regular. For example, the four influenza reports in May 2016 are shown in Table~\ref{table:poland_flu}.

\begin{table}[!ht]
    \centering
    \caption{\textbf{Influenza reporting intervals in Poland in May 2016.} Instead of reporting data at regular intervals (e.g., every 7 days), Poland reports data four ``weeks'' a month, regardless of the length of the month. This yields irregular interval durations. Here, the interval start and end are inclusive.}
    \label{table:poland_flu}
    \begin{tabular}{c | c | c | c}
        \textbf{Interval start} & \textbf{Interval end} & \textbf{Duration (days)} & \textbf{Source} \\
        \hline
        \texttt{2016-05-01} & \texttt{2016-05-07} & 7 & \cite{Poland5A2016} \\
        \texttt{2016-05-08} & \texttt{2016-05-15} & 8 & \cite{Poland5B2016} \\
        \texttt{2016-05-16} & \texttt{2016-05-22} & 7 & \cite{Poland5C2016} \\
        \texttt{2016-05-23} & \texttt{2016-05-31} & 9 & \cite{Poland5D2016}
    \end{tabular}
\end{table}

One remaining concern related to date and time is time zone. With the increasing use of internet data streams in disease forecasting and surveillance, it is important to be able to precisely place reference epidemiological data since associated internet data streams might be timestamped down to the second. Many data sources fail to report a time zone, so local time is often assumed. An incorrect time zone may impact analysis of high resolution data. For example, norovirus data are sometimes provided hourly (e.g., \cite{Guzman-Herrador2011, Mayet2011}), and time zone errors could have a potentially drastic negative effect on model results or analysis.

\paragraph*{Geography}

Political boundaries and names must be carefully managed. Subtle differences in names (e.g., Zurich vs. Z{\"u}rich) may lead to incorrect results during an analysis. The ISO 3166 standard defines country and principle subdivision (e.g., state or province) names, but it does not handle finer-than-subdivision regions, such as counties, districts, or cities.

Moreover, political boundaries (and thus populations and demographics) change over time. For example, South Sudan's split from Sudan in 2011 decreased Sudan's population by more than ten million people and dramatically changed its political boundary. Computing the historical attack rate for a disease (e.g., influenza incidence per 100,000 people), for instance, must take into account these changes.
\subsection{Data reporting challenges}

Beyond interface and data format challenges, there are challenges that lie within the bureaucratic \emph{reporting} process for an epidemiological institution. Modern disease surveillance systems rely on complex reporting hierarchies; raw data are initially captured at each provider, who then anonymizes and aggregates data as necessary before sending it to the next level in the hierarchy (perhaps a local or state public health department)~\cite{Fairchild2014a}. This hierarchy can have many levels. Even in many of the most developed regions of the world, much of this process continues to be done by hand, although the push to electronic medical records is gaining traction. As a result, most disease surveillance systems across the world experience reporting lags of at least one to two weeks.

This reporting lag can, in some cases, affect both an intuitive understanding of the situation as well as computational forecasting models. In an effort to combat surveillance system reporting lag, a number of attempts have been made to ``fill in'' the gaps using internet data (e.g., \cite{Polgreen2008, Ginsberg2009, Signorini2011, Generous2014}), but these studies require moderate to high levels of internet usage in the locales of interest, which are often not guaranteed.

Another issue is heterogeneous case definitions across jurisdictions. Many times, the case definitions used in epidemiological data are not clearly defined, and it is often difficult to navigate websites to identify the definitions. For example, many of the influenza surveillance systems in Europe use common, but not identical, case definitions~\cite{Aguilera2003}. Contextual differences in case definitions for Ebola~\cite{WHOEbola2017} could make interpreting data for the 2014 West African Ebola outbreak difficult.

One must also be concerned about language issues. Data are often provided in the native language of the region of the world in which they originate. For example, Thailand's Bureau of Epidemiology website~\cite{ThailandBoE2016} is natively displayed in Thai but also offers an English version~\cite{EnglishThailandBoE2016}. While online language translation services do exist (e.g., Google Translate\footnote{\url{https://translate.google.com/}}), these are not always reliable, and they cannot easily translate text in images (e.g., a website header comprised of images). To assist with language issues, formal disease- and epidemiological-focused ontologies (e.g., \cite{DiseaseOntology2011}) can help; translations, abbreviations, and alternate names can be encoded in an ontology to help automatically map different records to the same concept.

Furthermore, even within a single institution, there are often reporting nuances among diseases. For example, one context may be reported monthly, while another context may be reported weekly or daily. Some contexts may not be regularly reported; irregular reporting can lead to questions like, ``Is the value for a missing timestamp zero or unknown?''

Finally, case count data are often retroactively updated as new data are made available. In other words, historical data are not fixed the first time they are published. For example, a case count data point published today may be updated next week or the following week, as new data appear. This problem, often called ``backfill'', is due to the number and variety of members that comprise the complex reporting hierarchy that modern disease surveillance systems rely on; if a surveillance member's computer system goes down temporarily, for example, it may not be able to submit its data until the following week. Backfill can in some cases drastically affect analyses, so analysts and modelers must be aware of this potential issue~\cite{Farrow2017, Osthus2017a}.
\section{Conclusions}

We have identified three key challenges involving epidemiological data:
\begin{inparaenum}[1)]
    \item interface challenges,
    \item data format challenges, and
    \item data reporting challenges.
\end{inparaenum}
Each of these challenges can be addressed to simplify the efforts of analysts and modelers. Here, we propose a framework of best practices comprised of modern standards that should be adhered to when releasing epidemiological data to the public:

\begin{enumerate}
    \item Present the user with an interactive web interface to search and filter data. This interface should allow users to export data in common open formats (e.g., JSON using the EpiJSON standard, CSV).
    \item Provide a web-based API to allow automated data retrieval.
    \item Always use ISO 8601 dates, times, timestamps, and durations. Timestamps should either explicitly provide the local time zone or be adjusted for UTC, as specified by the ISO 8601 standard.
    \item When providing time series, clearly define the interval type so that timestamps can be interpreted properly.
    \item When possible, use ISO 3166 location names.
    \item Ensure all data are encoded using UTF-8.
    \item Ensure website can be run through an online language translation service (e.g., do not place important text in an image).
    \item When reporting case counts, the case definitions should be made explicit and clear.
    \item Clearly distinguish between unknown and zero values.
\end{enumerate}

These suggestions are not prescribing a single format or process; instead, these items provide a means for clearly defining and presenting epidemiological data to the public.

We implore the members of the global public health community to work together to create and follow standards for publishing data. Many institutions attempt to publish similar types of data using similar interfaces. In general, a user selects locations, diseases, optional time periods, and optional demographics in order to retrieve the desired data. Because many analysts and modelers have similar data desires, we feel this provides an opportunity for a generic shared epidemiological data access platform. Currently used by the CDC, one possibility might be Socrata\footnote{\url{https://socrata.com/}}, a platform that allows governments to share data openly. Socrata provides not only a modern interactive web interface but also an extensive API. Another option may be for public health institutions to collaboratively develop a free and open source solution that each could use. Such a platform may be more easily implemented by resource-constrained public health departments that have neither the time nor the money to develop their own solutions. Additionally, as web standards evolve, a shared data access platform could be updated in order to propagate these changes to each institution.

If a standard platform could be employed by institutions worldwide, then one could envision a future where global data could be easily collected without the challenges we currently face. This would in turn streamline epidemiological and public health analysis, modeling, and informatics, resulting in better public health decision-making capabilities.

Additionally, while this paper focuses on the public health and epidemiological communities, many of the challenges and solutions discussed here are not unique to them. Many of these same challenges are present whenever data of the same type are published globally by separate institutions that do not have an a priori agreed-upon set of standards. For example, weather and economic data have many of the same features as epidemiological data (e.g., locations, time intervals) and should also adhere to the ISO 8601 and 3166 standards, be encoded in UTF-8, and clearly distinguish between unknown and zero values.

Finally, it is important to recognize that this paper focuses on capable public health institutions with enough funding to collect and disseminate their epidemiological data. It should be noted, however, that a number of regions worldwide do not meet this criterion and are struggling even to monitor and care for their constituent populations, let alone publish reliable data. Unfortunately, it is precisely in these underserved regions that the public health community often desires data. Until worldwide public health infrastructure improves significantly, the suggestions here will remain peripheral for many; thus, the problems and suggested solutions put forth in this paper are likely only to be relevant to well-funded institutions. While the solutions presented in this paper may not be as effective at present time due to the lack of coverage, we offer them in preparation for the expanding global coverage that is continuously occurring, and it is only a matter of time until 100\% global internet coverage becomes a reality.
\section*{Funding}

This work was supported by the Defense Threat Reduction Agency's Joint Science and Technology Office for Chemical and Biological Defense under project numbers CB3656 and CB10007.
\section*{Acknowledgements}

This work was supported by the U.S. Department of Energy through the Los Alamos National Laboratory. Los Alamos National Laboratory is operated by Triad National Security, LLC, for the National Nuclear Security Administration of U.S. Department of Energy (Contract No. 89233218NCA000001).

\bibliographystyle{ieeetr}
\bibliography{references}

\end{document}